\documentclass[twocolumn,english,journal,final,twocolumn]{IEEEtran}
\usepackage[T1]{fontenc}
\usepackage[latin9]{inputenc}
\usepackage{color}
\usepackage{mathrsfs}
\usepackage{amsthm}
\usepackage{amsmath}
\usepackage{amssymb}
\usepackage{graphicx}
\usepackage{setspace}
\usepackage{esint}

\makeatletter
\theoremstyle{plain}
\newtheorem{thm}{\protect\theoremname}
\theoremstyle{plain}
\newtheorem{prop}[thm]{\protect\propositionname}

\IEEEoverridecommandlockouts

\usepackage{psfrag}\usepackage{bm}\usepackage{cite}\usepackage{babel}\usepackage{multirow}\usepackage{hhline}\usepackage{enumerate}\usepackage{amscd}\usepackage{calc}\usepackage{epsf}\usepackage{graphics}\usepackage{subfigure}

\makeatother

\makeatother

\usepackage{babel}
\makeatother

\usepackage{babel}
\providecommand{\propositionname}{Proposition}
\providecommand{\theoremname}{Theorem}

\begin{document}

\title{Energy-Efficient Adaptive Power Allocation for Incremental MIMO Systems\thanks{The authors are with Department of Electrical and Computer Engineering,
McGill University, Montréal, Canada. (E-mail: \{chaitanya.tumula@mail.mcgill.ca;
tho.le-ngoc@mail.mcgill.ca).}}

\author{Tumula V. K. Chaitanya, \emph{Member}, IEEE, and Tho Le-Ngoc, \emph{Fellow},
IEEE}
\maketitle
\begin{abstract}
We consider energy-efficient adaptive power allocation for three incremental
multiple-input multiple-output (IMIMO) systems employing ARQ, hybrid
ARQ (HARQ) with Chase combining (CC), and HARQ with incremental redundancy
(IR), to minimize their rate-outage probability (or equivalently packet
drop rate) under a constraint on average energy consumption per data
packet. We first provide the rate-outage probability expressions for
the three IMIMO systems, and then propose methods to convert them
into a tractable form and formulate the corresponding non-convex optimization
problems that can be solved by an interior-point algorithm for finding
a local optimum. To further reduce the solution complexity, using
an asymptotically equivalent approximation of the rate-outage probability
expressions, we approximate the non-convex optimization problems as
a unified geometric programming problem (GPP), for which we derive
the closed-form solution. Illustrative results indicate that the proposed
power allocation (PPA) offers significant gains in energy savings
as compared to the equal-power allocation (EPA), and the simple closed-form
GPP solution can provide closer performance to the exact method at
lower values of rate-outage probability, for the three IMIMO systems.\end{abstract}


\section{Introduction\label{sec:Introduction}}

Multiple-input multiple-output (MIMO) transmission schemes are most
suitable for systems with high spectral efficiency requirement. Despite
having many advantages, one of the fundamental limitations of MIMO
systems is the cost, increased power/energy consumption and the complexity
associated with their implementation in practical systems \cite{single_RF_MIMO}.
Towards addressing these problems associated with the conventional
MIMO systems, spatial modulation (SM) has been proposed in \cite{SM_1}
as a low-complexity MIMO transmission scheme that can improve the
energy efficiency (EE) with only channel state information known at
the receiver \cite{SM_2}. 

Incremental MIMO (IMIMO) \cite{IMIMO,IMIMO-2} is a variation of SM,
in which the multiple antennas at the transmitter are used in an incremental
fashion by utilizing the ARQ feedback to improve the reliability.
In an IMIMO system, the encoder functionality is simplified by letting
only one antenna from the transmit antenna array to be used to transmit
the information at any given time. Because of this, only a single
RF chain and a single power amplifier can be used on the transmitter
side and the receiver with multiple antennas can decode the message
optimally with relatively low complexity. After sending the information
from a chosen transmit antenna, the transmitter waits for the ARQ
feedback. If the information is received successfully, the receiver
sends a positive acknowledgment (ACK) and the next packet in the queue
is transmitted in the next transmission round. If the transmission
is not successful, a negative ACK (NACK) is sent from the receiver,
the same message is encoded and sent through a different transmit
antenna to exploit the spatial diversity. There are three possible
ways in which the encoding and decoding operations can be performed
during the transmission of an erroneous packet and they are IMIMO
using ARQ, CC-HARQ and IR-HARQ, respectively. Readers are encouraged
to refer to \cite{IMIMO,IMIMO-2} for more details of the three IMIMO
systems considered in this work and their advantages over the conventional
MIMO systems. 

\textbf{\emph{Related Work}}: Previous works on MIMO with ARQ considered
different aspects of the system performance. In \cite{MIMO_DMDT},
diversity-multiplexing-delay tradeoff of MIMO ARQ systems has been
studied. A multi-bit feedback scheme for MIMO IR-HARQ was proposed
and an outage analysis was presented in \cite{MIMO_Multibit}. A progressive
ARQ precoder design for MIMO transmission systems to minimize the
mean-square error has been proposed in \cite{MIMO_Precoder}. The
idea of using ARQ feedback for low-complexity MIMO system implementation
was proposed in \cite{IMIMO,IMIMO-2} along with an outage analysis
of IMIMO systems employing three retransmission mechanisms. \textcolor{black}{In
\cite{Makki}, among other things, the authors showed that for many
MIMO-ARQ schemes, the efficiency of ARQ protocols is dependent on
the considered scheme through the accumulated mutual information and
is independent of the performance metric. }

Recently many works have been focusing on the optimization of resources
in HARQ systems when the channel state information (CSI) is not available
at the transmitter. \textcolor{black}{A fixed outage probability analysis
of HARQ in block-fading channels with statistical CSI at the transmitter
was presented in \cite{Jindal}.} Optimal power allocation for improving
the average rate performance of HARQ schemes was presented in \cite{On_the_average_rate}
for quasi-static fading channels with different forms of CSI feedback.
Power adaptation to minimize the average transmission power under
a fixed rate-outage probability constraint for both the IR- and CC-HARQ
schemes was studied in \cite{Makki_TVT}. A rate allocation and adaptation
policy based on dynamic programming (DP) was proposed in \cite{Rate_adaptation_multi_bit}
for truncated IR-HARQ systems. In the works of \cite{HARQ_Paper,CC-HARQ-2},
the authors proposed power adaptation for IR- and CC-HARQ systems
in single-input single-output (SISO) i.i.d. Rayleigh fading channels
to minimize the rate-outage probability under an average energy constraint. 

\textbf{\emph{Contributions}}: We consider the problem of minimizing
the rate-outage probability of IMIMO systems employing ARQ, CC-HARQ
and IR-HARQ under a constraint on average energy consumption per packet.
In particular, we first generalize the system model of \cite{IMIMO}
to allow for adaptation of transmission power in different ARQ rounds
and provide the expressions for the rate-outage probability of IMIMO
employing ARQ, CC-HARQ and IR-HARQ. We then formulate the optimization
problems for each of the three IMIMO schemes using the derived outage
probability expressions. However, the given rate-outage probability
expressions are not mathematically tractable to be used in an optimization
problem formulation. Hence, we propose methods to convert these expressions
into a tractable form and formulate a non-convex optimization problem
that can be solved by an interior-point algorithm for finding a local
optimum. To further reduce the solution complexity, we propose an
asymptotically equivalent approximation of the derived rate-outage
probability expressions to approximate the non-convex optimization
problems as a unified geometric programming problem (GPP), for which
the closed-form solution is derived. 

Even though we consider the same optimization problem as in \cite{HARQ_Paper,CC-HARQ-2,CC-HARQ-3},
the present work differs in terms of the system model in the sense
that here we consider low-complexity IMIMO systems which utilize the
ARQ feedback to exploit the spatial diversity, whereas, in \cite{HARQ_Paper,CC-HARQ-2,CC-HARQ-3},
point-to-point SISO systems with IR and CC-HARQ were considered.

\section{System Model and Rate-Outage Analysis\label{sec:System-Model}}

\begin{figure}[tp]
\centering{}\textsf{\psfrag{TVK}[][][0.8][0]{Encoder}\psfrag{KVT}[][][0.7][0]{Decoder}\psfrag{1}[][][0.7][0]{$1$}\psfrag{2}[][][0.7][0]{$2$}\psfrag{M}[][][0.7][0]{$M$}\psfrag{N}[][][0.7][0]{$N$}\psfrag{ABC}[][][0.7][0]{ACK/NACK feedback}\psfrag{H}[][][0.7][0]{Channel}\includegraphics[clip,scale=0.31]{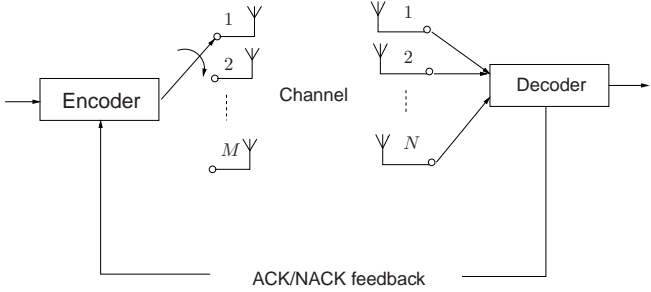}}\protect\caption{System model for the IMIMO.\label{fig:IMIMO_System_Model}}
\end{figure}
We consider a point-to-point IMIMO system having $M$ antennas at
the transmitter and $N$ antennas at the receiver as shown in Fig.
\ref{fig:IMIMO_System_Model}. We assume a frequency-flat Rayleigh
block-fading channel. The fading coefficient between the $m$th transmitting
antenna and the $n$th receiving antenna $h_{n,m},1\leq n\leq N\mbox{ and }1\leq m\leq M$
is i.i.d. with distribution $\mathcal{CN}\left(0,1\right)$. It is
assumed that $h_{n,m}$ remains unchanged during a fading block of
a fixed number of transmissions, and change independently from one
block to another. As in \cite{IMIMO,IMIMO-2}, we define one IMIMO
round as up to $L$ possible transmissions for each data packet. If
the destination is not able to decode a packet after $L$ transmission
attempts, the packet is dropped. The transmitter is assumed to have
only statistical knowledge of the fading coefficients, whereas, the
receiver is assumed to know the fading coefficients perfectly. Moreover,
in IMIMO systems, a new transmit antenna is used for sending an erroneous
packet in each transmission round (ARQ round). Hence the effective
channel changes independently \textcolor{black}{within} each ARQ round
of IMIMO and the instantaneous CSI feedback from the receiver is not
useful. We also assume that $L\leq M$ so that the quasi-static Rayleigh
block-fading IMIMO system model described above can be seen as a single-input
multiple-output (SIMO) system using ARQ or HARQ, in which the channel
fading block is equivalent to one ARQ round.

\textcolor{black}{Each ARQ round consists of $T$ symbols. We assume
that the modulation symbols have unit average energy and the same
power scaling factor $P_{l}$ is applied to all the $T$ symbols during
the $l$th ARQ round.} We write the received signal during the $t$th
channel use of $l$th ARQ round using the $m$th antenna for transmission
as $\mathbf{y}_{l}\left[t\right]=\sqrt{P_{l}}\mathbf{h}_{m}x_{m,l}\left[t\right]+\mathbf{z}_{l}\left[t\right]$,
where $\mathbf{h}_{m}=\left[h_{1,m},\ldots,h_{N,m}\right]^{T}$ denotes
the channel vector from the $m$th antenna to the $N$ antennas at
the receiver. The index of the antenna used for transmission in each
ARQ round is assumed to be known to the receiver. $x_{m,l}\left[t\right]\in\mathcal{C}$
denotes the modulation symbol from the $m$th transmit antenna in
the $t$th channel use of the $l$th ARQ round. $\mathbf{y}_{l}\left[t\right]=\left[y_{1,l}\left[t\right],\ldots,y_{N,l}\left[t\right]\right]^{T}$
denotes the channel output and $\mathbf{z}_{l}\left[t\right]=\left[z_{1,l}\left[t\right],\ldots,z_{N,l}\left[t\right]\right]^{T}$
represents the noise at the receiver and we assume $z_{n,l}\left[t\right]\sim\mathcal{CN}\left(0,1\right)$
for $n=1,\ldots,N$ and $l=1,\ldots,L$. The codebook construction
and decoding operations for each of the three IMIMO schemes has been
described in \cite{IMIMO,Gaussian_collison}. \textcolor{black}{Using
similar assumptions as in \cite{HARQ_Paper,CC-HARQ-2,CC-HARQ-3} about
the codewords, we consider rate-outage probability defined as the
probability that the instantaneous rate is smaller than the target
rate as a performance metric. }

\subsection{Rate-Outage Analysis of IMIMO Employing ARQ\label{sub:Outage_IMIMO-with-ARQ}}

For the case of IMIMO employing ARQ, the receiver only uses the information
from the current ARQ round to decode a message. For a target transmission
rate of $R$ bps/Hz, the probability of outage after $l$ ARQ rounds
is given by:\begin{subequations}
\begin{align}
\mathrm{p}_{\mathrm{out,}l}^{\mathrm{IMIMO,ARQ}} & \triangleq\prod_{k=1}^{l}\Pr\left\{ \log\left(1+P_{k}\left\Vert \mathbf{h}_{k}\right\Vert ^{2}\right)<R\right\} \label{eq:IMIMO_ARQ_Outage_1}\\
 & \!\!\!\!\!\!\!\!\!\!\!\!\!\!\!\!\!\!\!\!\!\!\!\!\!\!\!\!\!\!=\prod_{k=1}^{l}\gamma\left(N,Z_{k}\right)\label{eq:IMIMO_ARQ_Outage_2}\\
 & \!\!\!\!\!\!\!\!\!\!\!\!\!\!\!\!\!\!\!\!\!\!\!\!\!\!\!\!\!\!=\left(Z_{1}\cdots Z_{l}\right)^{N}\!e^{-\left(\sum_{k=1}^{l}Z_{k}\right)}\!\prod_{k=1}^{l}\sum_{n=0}^{\infty}\!\frac{\left(Z_{k}\right)^{n}}{\left(N\right)\!\ldots\left(N+n\right)}\label{eq:IMIMO_ARQ_Outage_3}\\
 & \!\!\!\!\!\!\!\!\!\!\!\!\!\!\!\!\!\!\!\!\!\!\!\!\!\!\!\!\!\!=\frac{\left(Z_{1}Z_{2}\cdots Z_{l}\right)^{N}}{N^{l}}+O\left(\frac{1}{P_{\mathrm{min}}^{lN+1}}\right)\label{eq:IMIMO_ARQ_Outage_4}
\end{align}
\end{subequations}where $\gamma\left(s,x\right)=\frac{1}{\Gamma\left(s\right)}\int_{0}^{x}t^{s-1}e^{-t}\mathrm{d}t$
is the normalized lower incomplete Gamma function, $\Gamma\left(N\right)$
is the Gamma function and $Z_{k}=\left(2^{R}-1\right)/P_{k},1\leq k\leq l$.
The relation in (\ref{eq:IMIMO_ARQ_Outage_2}) uses the fact that
$\left\Vert \mathbf{h}_{k}\right\Vert ^{2}\sim\chi^{2}$ distributed
random variable with $2N$ degrees of freedom and whose probability
density function is given by $f_{\left\Vert \mathbf{h}_{k}\right\Vert ^{2}}\left(\mathfrak{h}\right)=\frac{1}{\Gamma\left(N\right)}\mathfrak{h}^{N-1}e^{-\mathfrak{h}},\mathfrak{h}\geq0$.
In (\ref{eq:IMIMO_ARQ_Outage_4}), $P_{\mathrm{min}}=\min\left(P_{1},\cdots,P_{l}\right)$,
and we have written the rate-outage probability as the sum of the
first term and the higher-order terms.

\subsection{Rate-Outage Analysis of IMIMO Employing CC-HARQ\label{sub:Outage_IMIMO-with-CC-HARQ}}

In case of IMIMO employing CC-HARQ, the receiver combines the information
received across different transmission rounds using maximal-ratio-combining
(MRC). The rate-outage probability after $l$ ARQ rounds can be expressed
as:
\begin{align}
\mathrm{p}_{\mathrm{out,}l}^{\mathrm{IMIMO,CC-HARQ}} & \triangleq\Pr\left\{ \log\left(1+\sum_{k=1}^{l}P_{k}\left\Vert \mathbf{h}_{k}\right\Vert ^{2}\right)<R\right\} \nonumber \\
 & =\Pr\left\{ \sum_{k=1}^{l}\alpha_{k}<\underbrace{2^{R}-1}_{\triangleq Z}\right\} \label{eq:IMIMO_CC_HARQ_Outage_1}
\end{align}
where $\alpha_{k}\triangleq P_{k}\left\Vert \mathbf{h}_{k}\right\Vert ^{2},1\leq k\leq l$
has Gamma distribution with the shape parameter $N$ and the scale
parameter $P_{k}$. The term $\Theta\triangleq\sum_{k=1}^{l}\alpha_{k}$
is a sum of independent and non-identically distributed Gamma random
variables. Using the results from \cite{Aalo_1,kalyani_laurcella},
we can express (\ref{eq:IMIMO_CC_HARQ_Outage_1}) as:\footnote{\textcolor{black}{For a detailed derivation of the expressions, readers
can refer to \cite{Aalo_1,Aalo_2} and the references therein. }}\begin{subequations} 

\begin{align}
\mathrm{p}_{\mathrm{out,}l}^{\mathrm{IMIMO,CC-HARQ}}\nonumber \\
 & \!\!\!\!\!\!\!\!\!\!\!\!\!\!\!\!\!\!\!\!\!\!\!\!\!\!\!\!\!\!\!\!\!\!\!\!\!\!\!\!\!\!\!=\!\frac{1}{2}\!-\!\frac{1}{\pi}\!\int_{0}^{\infty}\!\frac{\sin\!\left(\!\sum_{k=1}^{l}N\tan^{-1}\!\left(xP_{k}\right)\!-\!Zx\right)}{\prod_{k=1}^{l}\left(1+\left(xP_{k}\right)^{2}\right)^{\frac{N}{2}}}\frac{dx}{x}\label{eq:IMIMO_CC_HARQ_Outage_2}\\
 & \!\!\!\!\!\!\!\!\!\!\!\!\!\!\!\!\!\!\!\!\!\!\!\!\!\!\!\!\!\!\!\!\!\!\!\!\!\!\!\!\!\!\!=\!\frac{\left(Z_{1}\cdots Z_{l}\right)^{N}}{\Gamma\left(lN+1\right)}\!\underbrace{\!\sum_{i_{1}=0}^{\infty}\!\!\cdots\!\!\sum_{i_{l}=0}^{\infty}\!\frac{\left(N\right)_{i_{1}}\!\cdots\left(N\right)_{i_{l}}\!}{\left(Nl+1\right)_{i_{1}+\cdots+i_{l}}}\left(\prod_{k=1}^{l}\frac{\left(\!-Z_{k}\!\right)^{i_{k}}}{i_{k}!}\right)}_{\triangleq\Phi_{2}^{l}\!\left(\!N,\!\cdots\!,\!N;Nl+1;\!-Z_{1},\!\cdots\!,\!-Z_{l}\!\right)}\label{eq:IMIMO_CC_HARQ_Outage_3}\\
 & \!\!\!\!\!\!\!\!\!\!\!\!\!\!\!\!\!\!\!\!\!\!\!\!\!\!\!\!\!\!\!\!\!\!\!\!\!\!\!\!\!\!\!=\frac{\left(Z_{1}Z_{2}\cdots Z_{l}\right)^{N}}{\Gamma\left(lN+1\right)}\!+\!O\!\left(\!\frac{1}{P_{\mathrm{min}}^{lN+1}}\!\right)\label{eq:IMIMO_CC_HARQ_Outage_4}
\end{align}
\end{subequations}where the notation $\left(x_{1}\right)_{y_{1}}=\Gamma\left(x_{1}+y_{1}\right)/\Gamma\left(x_{1}\right)$
with $\left(x_{1}\right)_{0}=1$ and the function $\Phi_{2}^{l}\left(.\right)$
in (\ref{eq:IMIMO_CC_HARQ_Outage_3}) is the confluent Lauricella
hypergeometric function of $l$ variables \cite{Exton}. In (\ref{eq:IMIMO_CC_HARQ_Outage_4}),
we have written the rate-outage probability expression as the sum
of the first term and the higher-order terms.

\subsection{Rate-Outage Analysis of IMIMO Employing IR-HARQ\label{sub:Rate-Outage-Analysis-of-IR-HARQ}}

The rate-outage probability after $l$ ARQ rounds for an IMIMO system
with IR-HARQ can be expressed as \cite{IMIMO,IMIMO-2}:
\begin{eqnarray}
\mathrm{p}_{\mathrm{out,}l}^{\mathrm{IMIMO,IR-HARQ}} & \triangleq & \Pr\left\{ \sum_{k=1}^{l}\log\left(1+P_{k}\left\Vert \mathbf{h}_{k}\right\Vert ^{2}\right)<R\right\} \nonumber \\
 &  & =2^{R}{\color{blue}{\color{black}g}}_{l}\left(-R\ln2\right)-{\color{blue}{\color{black}g}}_{l}\left(0\right)\label{eq:IMIMO_IR_HARQ_Outage_1}
\end{eqnarray}
where,
\begin{align*}
{\color{blue}{\color{black}g}}_{l}\left(t\right) & =q_{1}\left(t\right)*q_{2}\left(t\right)*\cdots*q_{l}\left(t\right),\\
q_{1}\left(t\right) & =-\!e^{t}\!\left(\!1\!-\!\gamma\!\left(\!N,\frac{e^{-t}-1}{P_{1}}\!\right)\!\right)\!u\left(-t\right)\!-\!e^{t}\!\left(1\!-\!u\left(-t\right)\right),\mbox{ and }\\
q_{i}\left(t\right) & =\frac{\left(e^{-t}\!-1\right)^{N-1}e^{t+\frac{1-e^{-t}}{P_{i}}}}{P_{i}^{N}\Gamma\left(N\right)}u\left(-t\right),2\leq i\leq l
\end{align*}
with $u\left(t\right)$ being the unit step function defined as $u\left(t\right)=1,\mbox{ for }t>0\mbox{ and }u\left(t\right)=0,\mbox{\,\ for\,\,}t<0$,
and the symbol $*$ represents the convolution operation. The derivation
of $\mathrm{p}_{\mathrm{out,}l}^{\mathrm{IMIMO,IR-HARQ}}$ and $q_{1}\left(t\right)$
is similar to the derivations given in \cite{IMIMO}.\footnote{The second term in the equation (18) of \cite{IMIMO} should be $-e^{t}\left(1-u\left(-t\right)\right)$
instead of $-e^{-t}u\left(t-1\right)$.} We provide the derivation of $q_{i}\left(t\right),2\leq i\leq l$
in Appendix A. 

We use the Jensen's inequality in (\ref{eq:IMIMO_IR_HARQ_Outage_1})
together with the results from Section \ref{sub:Outage_IMIMO-with-CC-HARQ}
to write:
\begin{align}
\mathrm{p}_{\mathrm{out,}l}^{\mathrm{IMIMO,IR-HARQ}} & \geq\Pr\left\{ \log\left(1+\frac{1}{l}\sum_{k=1}^{l}P_{k}\left\Vert \mathbf{h}_{k}\right\Vert ^{2}\right)<\frac{R}{l}\right\} \nonumber \\
 & =\frac{\left(Z_{1}^{'}Z_{2}^{'}\cdots Z_{l}^{'}\right)^{N}}{\Gamma\left(lN+1\right)}+O\left(\frac{1}{P_{\mathrm{min}}^{lN+1}}\right)\label{eq:IMIMO_IR_HARQ_approx}
\end{align}
where $Z_{i}^{'}\triangleq Y\left(l\right)/P_{i},1\leq i\leq l\mbox{ and }Y\left(l\right)=l\left(2^{R/l}-1\right)$.

\section{Optimization Problems and Solutions\label{sec:Optimization-Problems-and-solutions}}

In this section, we first state the general optimization problem and
describe methods to solve the problem for each of three IMIMO systems
considered in this work. We define the average transmit energy per
packet as: 
\[
E_{\mathrm{avg}}\triangleq T\sum_{l=1}^{L}P_{l}\mathrm{p}_{\mathrm{out,}l-1}^{\mathrm{IMIMO,ARQ/HARQ\,type}}.
\]
We also define the quantity $\overline{E}_{\mathrm{avg}}\triangleq E_{\mathrm{avg}}/T$
for mathematical tractability. \textcolor{black}{Similar to \cite{HARQ_Paper,CC-HARQ-2},
we formulate the general optimization problem as:}
\begin{align}
 & \min_{\left(P_{1},P_{2},\ldots,P_{L}\right)}\mathrm{\mathrm{p_{out,\mathit{L}}^{IMIMO,ARQ/HARQ\,type}}}\nonumber \\
\mathrm{subject\,to\quad} & 0\leq P_{l},\quad\mbox{for}\quad1\leq l\leq L,\label{eq:Opt_prob_1}\\
 & \sum_{l=1}^{L}P_{l}\mathrm{p}_{\mathrm{out,}l-1}^{\mathrm{IMIMO,ARQ/HARQ\,type}}\leq\overline{E}_{\mathrm{given}}\nonumber 
\end{align}

\subsection{Solution for IMIMO employing ARQ\label{sub:Solution-for-IMIMO-ARQ}}

The rate-outage probability expressions for an IMIMO system employing
ARQ given in (\ref{eq:IMIMO_ARQ_Outage_2}) and (\ref{eq:IMIMO_ARQ_Outage_3})
involve product of integrals and an infinite summation, respectively.
Hence, for mathematical tractability, and to be used in (\ref{eq:Opt_prob_1}),
we approximate (\ref{eq:IMIMO_ARQ_Outage_2}) using the standard Gauss-Legendre
approximation as:
\begin{align}
\mathrm{p_{out,\mathit{l}}^{IMIMO,ARQ}} & =\frac{\left(Z_{1}\cdots Z_{l}\right)^{N}}{\left(\Gamma\left(N\right)\right)^{l}}\left(\prod_{k=1}^{l}\int_{0}^{1}\underbrace{t^{N-1}e^{-tZ_{k}}}_{=f_{k}}dt\right)\nonumber \\
 & \approx\!\frac{\left(Z_{1}\!\cdots Z_{l}\right)^{N}}{\left(2\,\Gamma\left(N\right)\right)^{l}}\!\left(\!\prod_{k=1}^{l}\sum_{i=1}^{M_{1}}w_{i}f_{k}\left(\!\frac{t_{i}+1}{2}\!\right)\!\right)\label{eq:IMIMO_ARQ_approx}
\end{align}
where $w_{i}$ and $t_{i}$ are, respectively, the $i$th weight and
the $i$th zero of the Legendre polynomial of order $M_{1}$ \cite[eq. (25.4.30)]{Handbook}.
\textcolor{black}{Note that the accuracy of the approximation in (\ref{eq:IMIMO_ARQ_approx})
depends on $M_{1}$.} An arbitrarily accurate approximation can be
obtained by selecting an appropriate value of $M_{1}$. In practical
systems using retransmission schemes with a typical value of $L=3$,
outage probability values in the order up to $10^{-4}$ are of interest
\cite{Comm_mag}, and \textcolor{black}{we observed through numerical
results}\footnote{The actual approximation error depends on the $2M_{1}$th derivative
of $f_{k}\left(t\right)$ \cite[eq. (25.4.30)]{Handbook}. Nonetheless
a reference value can be computed numerically by generating many random
variables and computing the rate-outage probability using $\prod_{k=1}^{l}\mathrm{Pr}(\left\Vert \mathbf{h}_{k}\right\Vert ^{2}<Z_{k})$.} that $M_{1}=1024$ approximates the outage probability values with
an approximation error smaller than $10^{-6}$. Using the approximation
in (\ref{eq:IMIMO_ARQ_approx}), we can write the optimization problem
in (\ref{eq:Opt_prob_1}) for ARQ as: 
\begin{align}
 & \!\!\!\!\!\!\!\!\!\!\!\!\!\!\!\!\!\!\!\!\!\!\!\min_{\left(P_{1},P_{2},\ldots,P_{L}\right)}\frac{\left(Z_{1}\cdots Z_{l}\right)^{N}}{\left(2\,\Gamma\left(N\right)\right)^{L}}\left(\prod_{k=1}^{L}\left[\sum_{i=1}^{M_{1}}w_{i}f_{k}\left(\frac{t_{i}+1}{2}\right)\right]\right)\nonumber \\
\mathrm{subject\,to\quad} & 0\leq P_{l},\mbox{ for }1\leq l\leq L,\label{eq:Opt_prob_IMIMO_ARQ-1}\\
 & \!\!\!\!\!\!\!\!\!\!\!\!\!\!\!\!\!\!\!\!\!\!\!\!\!\!\!\!\!P_{1}\!+\!\sum_{l=2}^{L}\!P_{l}\!\frac{\left(Z_{1}\!\cdots Z_{l}\right)^{N}}{\left(2\,\Gamma\left(N\right)\right)^{l-1}}\!\left(\!\prod_{k=1}^{l-1}\left[\!\sum_{i=1}^{M_{1}}w_{i}f_{k}\!\left(\!\frac{t_{i}+1}{2}\!\right)\!\right]\!\right)\!\leq\!\overline{E}_{\mathrm{given}}\nonumber 
\end{align}

The optimization problem in (\ref{eq:Opt_prob_IMIMO_ARQ-1}) is non-convex
and hence we are not guaranteed to find the globally optimum solution
to the problem unless an exhaustive search is performed. However,
nonlinear optimization techniques can be used to find a local optimum
of (\ref{eq:Opt_prob_IMIMO_ARQ-1}). We use an interior-point algorithm
outlined in \cite{Interior_point_algo} which uses either a Newton
step or a conjugate gradient step using a trust region to find a solution.
For each feasible point $\left(P_{1},\ldots,P_{L}\right)$, we need
to perform $\left(M_{1}L-M_{1}\right)$ function evaluations at the
zeros of the Legendre polynomial, hence the complexity of finding
a solution is high.

\subsection{Solution for IMIMO employing CC-HARQ\label{sub:Solution-for-IMIMO_CC-HARQ}}

The expressions given in (\ref{eq:IMIMO_CC_HARQ_Outage_2}) and (\ref{eq:IMIMO_CC_HARQ_Outage_3})
are not mathematically tractable as functions of optimization variables
$\left(P_{1},\cdots,P_{L}\right)$. Hence for mathematical tractability,
we approximate (\ref{eq:IMIMO_CC_HARQ_Outage_2}) using the standard
Gauss-Legendre approximation as:
\begin{align}
\mathrm{p}_{\mathrm{out},\mathit{l}}^{\mathrm{IMIMO,CC-HARQ}}\nonumber \\
 & \!\!\!\!\!\!\!\!\!\!\!\!\!\!\!\!\!\!\!\!\!\!\!\!\!\!\!\!\!\!\!\!\!\!\!\!\!\!\!\!\!\!\!\!\!=\frac{1}{2}\!-\!\frac{1}{\pi}\!\int_{0}^{1}\!\underbrace{\left(\!\!\frac{\sin\!\left(\!\sum_{k=1}^{l}N\tan^{-1}\!\left(\frac{t}{1-t}P_{k}\right)-\frac{Zt}{1-t}\right)}{\left[\!\prod_{k=1}^{l}\!\left(\!1+\left(\frac{t}{1-t}P_{k}\right)^{2}\!\right)^{\frac{N}{2}}\!\right]\!\left(t-t^{2}\right)}\!\!\right)}_{\triangleq k_{l}\left(t\right)}\!dt\nonumber \\
 & \!\!\!\!\!\!\!\!\!\!\!\!\!\!\!\!\!\!\!\!\!\!\!\!\!\!\!\!\!\!\!\!\!\!\!\!\!\!\!\!\!\!\!\!\!\approx\!\frac{1}{2}\!-\!\frac{1}{2\pi}\!\left[\!\sum_{i=1}^{M_{2}}w_{i}k_{l}\left(\frac{t_{i}+1}{2}\right)\!\right]\label{eq:Guass_legendre_approx}
\end{align}
In this case also, an arbitrarily accurate approximation can be obtained
by selecting an appropriate value of $M_{2}$. \textcolor{black}{We}
observed through numerical results that $M_{2}=512$ approximates
the outage probability values with an approximation error smaller
than $10^{-6}$. Using (\ref{eq:Guass_legendre_approx}), the optimization
problem in (\ref{eq:Opt_prob_1}) for CC-HARQ case can equivalently
be written as: 
\begin{alignat}{1}
 & \min_{\left(P_{1},P_{2},\ldots,P_{L}\right)}\frac{1}{2}-\frac{1}{2\pi}\left[\sum_{i=1}^{M_{2}}w_{i}k_{L}\left(\frac{t_{i}+1}{2}\right)\right]\nonumber \\
\mathrm{subject\,to\quad} & 0\leq P_{l},\mbox{ for }1\leq l\leq L,\label{eq:Opt_prob_IMIMO_CC_HARQ_1}\\
 & \!\!\!\!\!\!\!\!\!\!\!\!\!\!\!\!\!\!\!\!\!\!\!\!\!\!\!P_{1}+\sum_{l=2}^{L}P_{l}\left(\frac{1}{2}-\frac{1}{2\pi}\left[\sum_{i=1}^{M_{2}}w_{i}k_{l-1}\left(\frac{t_{i}+1}{2}\right)\right]\right)\leq\overline{E}_{\mathrm{given}}\nonumber 
\end{alignat}
We used the same interior-point algorithm outlined \cite{Interior_point_algo}
to find a solution for (\ref{eq:Opt_prob_IMIMO_CC_HARQ_1}).

\subsection{Solution for IMIMO employing IR-HARQ}

Using (\ref{eq:IMIMO_IR_HARQ_Outage_1}), the optimization problem
for the IR-HARQ case can be written as:
\begin{align}
 & \min_{\left(P_{1},P_{2},\ldots,P_{L}\right)}2^{R}{\color{blue}{\color{black}g}}_{L}\left(-R\ln2\right)-{\color{blue}{\color{black}g}}_{L}\left(0\right)\nonumber \\
\mathrm{subject\,to\quad} & 0\leq P_{l},\quad\mbox{for}\quad1\leq l\leq L,\label{eq:Opt_prob_IMIMO_IR_HARQ}\\
 & \!\!\!\!\!\!\!\!\!\!\!\!\!\!\!\!\!\!\!\!\!\!\!\!\!\!\!P_{1}+\sum_{l=2}^{L}P_{l}\left(2^{R}{\color{blue}{\color{black}g}}_{l-1}\left(-R\ln2\right)-{\color{blue}{\color{black}g}}_{l-1}\left(0\right)\right)\leq\overline{E}_{\mathrm{given}}\nonumber 
\end{align}
where the function ${\color{blue}{\color{black}g}}_{l}\left(t\right),1\leq l\leq L$
is obtained by the convolution of the functions $q_{1}\left(t\right),\cdots,q_{l}\left(t\right)$
defined in Section \ref{sub:Rate-Outage-Analysis-of-IR-HARQ}. We
can express this convolution operation in terms of a multiple-integral
in $l-1$ dimensions. Similar to the techniques used in Sections \ref{sub:Solution-for-IMIMO-ARQ}
and \ref{sub:Solution-for-IMIMO_CC-HARQ}, we can approximate the
finite dimensional integrals as finite sums using the Gauss-Legendre
approximation, or by applying the method described in \cite{2D} for
two dimensions. These finite summations can then be used in the optimization
problem of (\ref{eq:Opt_prob_IMIMO_IR_HARQ}) and can be solved using
interior-point methods.

\section{GPP Approach and Closed-form Solution\label{sec:GPP-Approach}}

In this section, we provide approximate expressions for the rate-outage
probability of IMIMO systems employing ARQ, CC-HARQ and IR-HARQ to
formulate an unified geometric programming problem (GPP), for which
the closed-form solution is derived. From the rate-outage probability
expressions for IMIMO using ARQ, CC-HARQ and IR-HARQ given in (\ref{eq:IMIMO_ARQ_Outage_4}),
(\ref{eq:IMIMO_CC_HARQ_Outage_4}) and (\ref{eq:IMIMO_IR_HARQ_approx}),
respectively, we neglect the higher-order terms and write an asymptotically
equivalent approximation as:
\begin{equation}
\mathrm{p}_{\mathrm{out,}l}^{\mathrm{IMIMO,ARQ/HARQ\,type}}\backsimeq\frac{W_{l}}{\left(P_{1}^{N}\cdots P_{l}^{N}\right)}.\label{eq:IMIMO_ARQ_Outage_approx}
\end{equation}
where
\begin{equation}
W_{l}=\begin{cases}
\frac{Z^{lN}}{N^{l}} & \mbox{for\,\ IMIMO using ARQ}\\
\frac{Z^{lN}}{\Gamma\left(lN+1\right)} & \mbox{for\,\ IMIMO using CC-HARQ}\\
\frac{Y\left(l\right)^{lN}}{\Gamma\left(lN+1\right)} & \mbox{for\,\ IMIMO using IR-HARQ}
\end{cases}\label{eq:A_l_def}
\end{equation}
The motivation for the approximation in (\ref{eq:IMIMO_ARQ_Outage_approx})
is: i) as $P_{\mathrm{min}}=\min\left(P_{1},\cdots,P_{l}\right)\rightarrow\infty$
, the $O\left(.\right)$ terms in (\ref{eq:IMIMO_ARQ_Outage_4}),
(\ref{eq:IMIMO_CC_HARQ_Outage_4}) and (\ref{eq:IMIMO_IR_HARQ_approx})
go to zero faster than the approximated terms in (\ref{eq:IMIMO_ARQ_Outage_approx}),
(\ref{eq:A_l_def}); and ii) the maximum possible diversity order
achievable in a Rayleigh fading channel for IMIMO system with $N$
receiving antennas after $l$ ARQ rounds is $lN$, which is also achieved
by the approximations in (\ref{eq:IMIMO_ARQ_Outage_approx}) and (\ref{eq:A_l_def}).
\textcolor{black}{Even though the asymptotically equivalent approximations
of the rate-outage probability for the three IMIMO methods have a
similar structure, they differ in terms of the coefficient $W_{l}$
as shown in (\ref{eq:A_l_def}). The similarity in the structure of
approximated outage probability expressions in (\ref{eq:IMIMO_ARQ_Outage_approx})
allows us to approximate the optimization problem in (\ref{eq:Opt_prob_1})
as a unified GPP as:}
\begin{align}
 & \min_{\left(P_{1},P_{2},\ldots,P_{L}\right)}\frac{W_{L}}{\left(P_{1}^{N}\cdots P_{L}^{N}\right)}\nonumber \\
\mathrm{Subject\,to\quad} & 0\leq P_{l},\quad\mbox{for}\quad1\leq l\leq L,\label{eq:Opt_prob_IMIMO_GPP}\\
 & P_{1}+\sum_{l=2}^{L}P_{l}\frac{W_{l-1}}{\left(P_{1}^{N}\cdots P_{l-1}^{N}\right)}\leq\overline{E}_{\mathrm{given}}\nonumber 
\end{align}
In the following theorem, we provide the closed-form solution for
the problem in (\ref{eq:Opt_prob_IMIMO_GPP}).
\begin{thm}
The closed-form solution for the problem in (\ref{eq:Opt_prob_IMIMO_GPP})
is given by:
\begin{align}
P_{1}^{*} & =\frac{\overline{E}_{\mathrm{given}}N\left(N+1\right)^{L-1}}{\left(N+1\right)^{L}-1},\nonumber \\
P_{i}^{*} & =\frac{W_{i-2}}{W_{i-1}\left(1+N\right)}\left(P_{i-1}^{*}\right)^{N+1},\,\mbox{ for }2\leq i\leq L\label{eq:Optimal_P_1}
\end{align}
\end{thm}
\begin{IEEEproof}
Please see Appendix B.
\end{IEEEproof}
\textcolor{black}{As can be seen from (\ref{eq:Optimal_P_1}), the
solution of the GPP approach differs for different IMIMO methods through
the coefficient values $W_{l},1\leq l\leq L$.}

\section{Numerical Illustrations and Discussion\label{sec:Numerical-Results}}

In this section, we present illustrative examples for a performance
comparison of the proposed power allocation (PPA) and equal power
allocation (EPA). For the non-convex optimization case, we solve the
optimization problems (\ref{eq:Opt_prob_IMIMO_ARQ-1}), (\ref{eq:Opt_prob_IMIMO_CC_HARQ_1})
and (\ref{eq:Opt_prob_IMIMO_IR_HARQ}) using the interior-point algorithm
presented in \cite{Interior_point_algo}; this method is labeled as
`PPA-exact method' in the plots.\footnote{Although we cannot guarantee the global optimum, we have verified
that the solution offered by the interior-point algorithm matches
very closely with the optimal solution found by an exhaustive grid
search over the possible values of $P_{l},1\leq l\leq L$.} For the `PPA-GPP approach' we solved the approximated optimization
problem in (\ref{eq:Opt_prob_IMIMO_GPP}) for IMIMO with ARQ, CC-HARQ
and IR-HARQ, respectively. For the EPA case, we solved for $P$ using
the additional constraint $P_{1}=\cdots=P_{L}=P,$ in (\ref{eq:Opt_prob_IMIMO_ARQ-1}),
(\ref{eq:Opt_prob_IMIMO_CC_HARQ_1}) and (\ref{eq:Opt_prob_IMIMO_IR_HARQ}).
\begin{figure}
\subfigure[IMIMO  with ARQ]{\label{fig:ARQ}\textsf{\includegraphics[clip,scale=0.47]{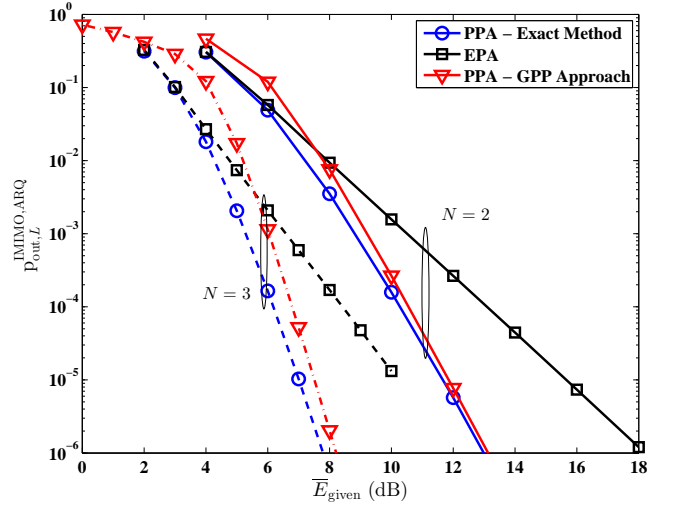}}}\quad{}\subfigure[IMIMO with CC-HARQ]{\label{fig:CC_HARQ}\textsf{\includegraphics[clip,scale=0.47]{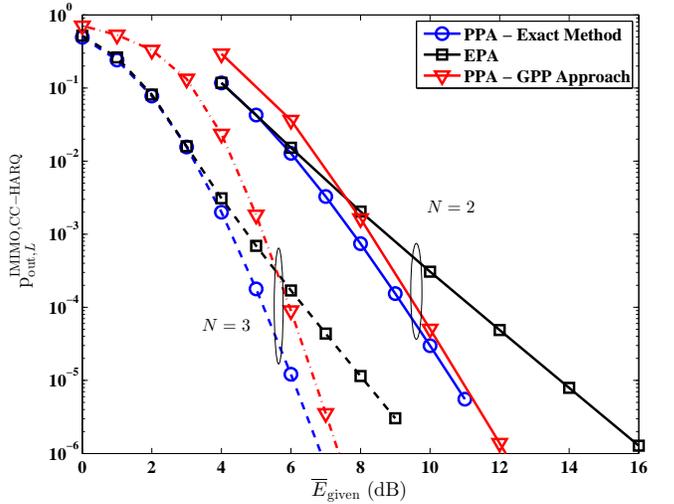}}}\hfill{}\subfigure[IMIMO with IR-HARQ]{\label{fig:IR_HARQ}

\textsf{\includegraphics[clip,scale=0.47]{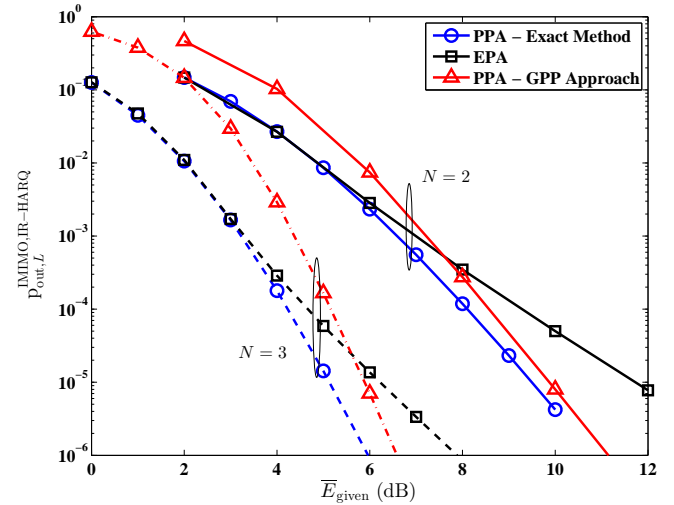}}}\hfill{}

\protect\caption{Performance comparison of the proposed power allocation with the equal
power allocation for an IMIMO system employing ARQ, CC-HARQ and IR-HARQ.
The parameters for the simulation are $L=2,M=2$ and $R=2$ bps/Hz.\label{fig:Perf_comp}}
\end{figure}

\textbf{\emph{Proposed Power Allocation vs Equal Power Allocation}}:
Figure \ref{fig:Perf_comp} shows a performance comparison of PPA
and EPA for IMIMO employing ARQ, CC-HARQ and IR-HARQ under different
system parameter values. We plotted $\mathrm{p}_{\mathrm{out,}l}^{\mathrm{IMIMO,ARQ/HARQ\,type}}$
as a function of $\overline{E}_{\mathrm{given}}$. Following observations
can be made from Figs. \ref{fig:ARQ}-\ref{fig:IR_HARQ}. First, for
higher values of outage probability, the EPA has a similar performance
as that of the \textquoteleft PPA-exact method\textquoteright , especially
when the diversity order is high. The gains offered by the `PPA-exact
method' over EPA are more significant for smaller values of rate-outage
probability (equivalently for higher average energy limit). Second,
for the case of $L=N=2$, at a rate-outage probability of $10^{-5}$,
the gain for the `PPA-exact method' over the the EPA solution is 4
dB, 3.1 dB and 2.3 dB, for IMIMO with ARQ, CC-HARQ and IR-HARQ, respectively.
However, the gains reduce as the diversity order of the system increases
(i.e., as the value of $N$ increases)\footnote{Note that one can increase the diversity order by increasing the value
of $L$ as well. However because of space constraints, we could not
show the results here.}. Third, in general, the closed-form `PPA-GPP approach\textquoteright{}
provides higher outage probability than the `PPA-exact method\textquoteright{}
and the performance gap between the two proposed schemes get closer
as the energy limit $\overline{E}_{\mathrm{given}}$ increases, especially
for smaller values of $L$ and $N$. \textcolor{black}{The reason
for the ``loss'' seen by the `PPA-GPP approach' relative to the
EPA for higher values of rate-outage probability is as follows. The
approximation error of outage probability expressions is non-negligible
for smaller values of $\overline{E}_{\mathrm{given}}$. The approximations
become tighter (asymptotically equivalent) and the performance of
the `GPP approach' matches that of the exact method as $\overline{E}_{\mathrm{given}}$
value increases. To reduce the loss of the `PPA-GPP approach' relative
to `PPA-exact method', one method is to use tighter approximations
by considering the higher-order terms of the rate-outage probability
expressions. However, when higher-order terms are also considered,
they may include both positive and negative terms in the approximations,
and this may restrict the use of the geometric programming approach
to find a solution. }
\begin{figure}
\textsf{\includegraphics[clip,scale=0.47]{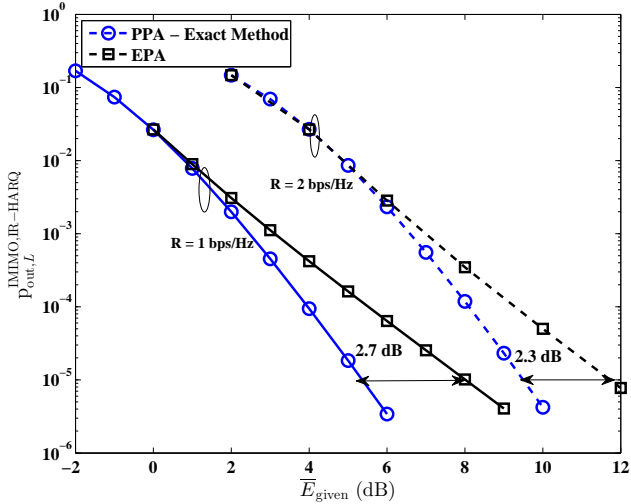}}

\protect\caption{Performance comparison of PPA-exact method and EPA for IMIMO system
employing IR-HARQ with different values of spectral efficiency, and
$M=L=N=2$. \label{fig:diff_R_comp}}
\end{figure}

\textbf{\emph{Comparison for Different Values of $R$}}: For an IMIMO
system employing ARQ and CC-HARQ, we have the following proposition.\setcounter{thm}{0}
\begin{prop}
In an IMIMO system employing ARQ and CC-HARQ, for a given maximum
number of transmissions $L$ and target rate-outage probability value
of $\rho$, if $\left(P_{1,R_{1}},P_{2,R_{1}},\cdots,P_{L,R_{1}}\right)$
is the optimal power allocation solution with the average energy $E_{\mathrm{avg},1}$
for a spectral efficiency $R_{1}>0$ , then the optimal power allocation
solutions and the average energy for a spectral efficiency of $R_{2}\neq R_{1}$
are given by
\begin{align}
P_{l,R_{2}} & =P_{l,R_{1}}\left(\frac{2^{R_{2}}-1}{2^{R_{1}}-1}\right),1\leq l\leq L,\label{eq:scale_optimum_power}\\
E_{\mathrm{avg},2} & =E_{\mathrm{avg},1}\left(\frac{2^{R_{2}}-1}{2^{R_{1}}-1}\right)\nonumber 
\end{align}
\end{prop}
\begin{IEEEproof}
The proof follows the same arguments as in the proof of Proposition
1 in \cite{CC-HARQ-2}.
\end{IEEEproof}
\setcounter{thm}{1}

\textcolor{black}{Hence, for an IMIMO system employing ARQ and CC-HARQ,
it is sufficient to solve the optimization problem in (\ref{eq:Opt_prob_1})
for a single value of $R$ and scale the resulting power values according
to (\ref{eq:scale_optimum_power}) to obtain the optimal power values
for a different value of $R$. In fact, a similar result as in Proposition
1 applies to EPA and GPP approaches as well. Hence for a given change
in the value of $R$, the performance of all power allocation methods
shift by the same amount and hence the relative performance difference
remains the same independent of the value of $R$. However, for an
IMIMO system with IR-HARQ, optimal power values for different values
of $R$ does not scale according to (\ref{eq:scale_optimum_power}),
and hence performance difference between `PPA-exact method' and EPA
is different for different values of $R$, this can be seen from Fig.
\ref{fig:diff_R_comp}.}
\begin{figure}
\subfigure[$L=2, M =2, N = 2, R = 2\, \mbox{bpcu}$.]{\label{fig:Power_1}\textsf{\includegraphics[clip,scale=0.47]{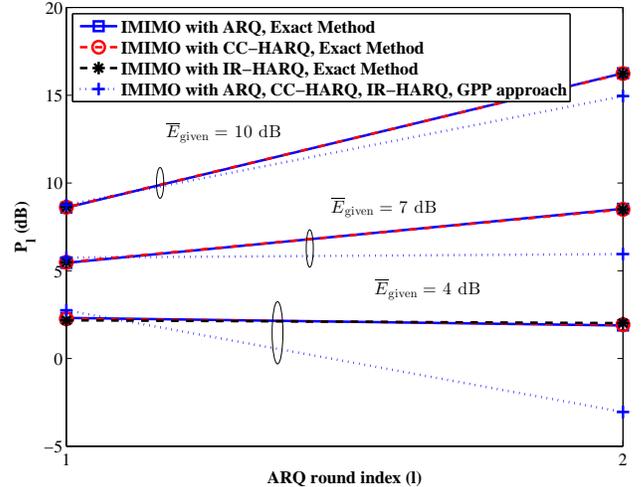}}}\hfill{}\\
\subfigure[$L=3, M= 3, N = 2, R = 2\, \mbox{bpcu}$.]{\label{fig:Power_2}\textsf{\includegraphics[clip,scale=0.47]{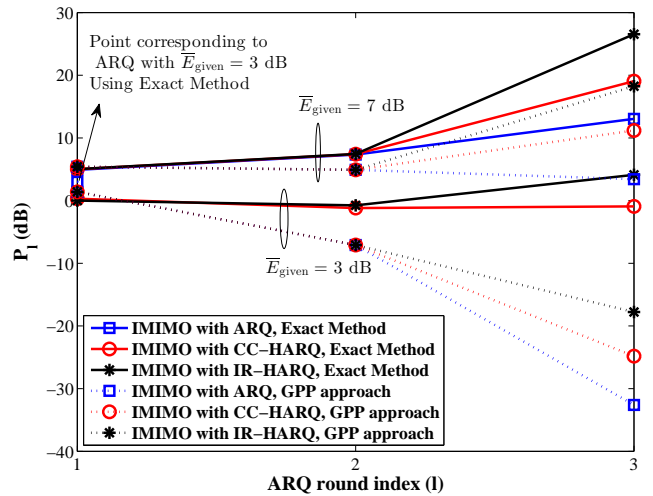}}}\hfill{}

\protect\caption{Comparison of $P_{l},1\leq l\leq L$ values to be used in different
ARQ rounds for the three IMIMO schemes for a given $\overline{E}_{\mathrm{given}}$.\label{fig:Power_comp}}
\end{figure}

\textbf{\emph{Power Values}}: Figure \ref{fig:Power_comp} shows the
power values obtained by solving the optimization problems using different
approaches.\textcolor{black}{{} As can be seen from Fig. \ref{fig:Power_1},
for a given value of $\overline{E}_{\mathrm{given}}$, and for $L=2$,
the three IMIMO systems have }\textcolor{black}{\emph{similar optimal
power values}}\textcolor{black}{{} obtained using the exact method,
and }\textcolor{black}{\emph{the same}}\textcolor{black}{{} power values
obtained using the GPP approach. The reason for this can be explained
by noting that, all the three optimization problems have the same
average energy constraint as $\mathrm{p}_{\mathrm{out},1}^{\mathrm{IMIMO,ARQ/HARQ\,type}}$
is the same for all the three IMIMO systems (both the exact and approximated
value). The approximated expression for packet drop probability $\mathrm{p}_{\mathrm{out},2}^{\mathrm{IMIMO,ARQ/HARQ\,type}}\propto1/\left(P_{1}^{2}P_{2}^{2}\right)$
for all the three IMIMO systems, and hence they have the same solution
using the GPP approach and similar optimal power values using the
exact method.} For $L=3$ in Fig. \ref{fig:Power_2}, we can clearly
see the difference in the values of $P_{l}$ for the three IMIMO schemes
using the exact method. Furthermore, as the value of $\overline{E}_{\mathrm{given}}$
increases, more power is allocated for ``later ARQ rounds'', which
are towards the end of the ARQ process. Since the objective of the
optimization is to minimize the outage after $L$ ARQ rounds, we need
to improve the probability of successful decoding during these later
ARQ rounds. In other words, the energy cost associated with an unsuccessful
decoding during these later ARQ rounds increases. Hence, for large
values of $\overline{E}_{\mathrm{given}}$, PPA assigns high power
values to these later ARQ rounds. We can also note from Fig. \ref{fig:Power_2}
that for smaller values of $\overline{E}_{\mathrm{given}}$, in case
of IMIMO with ARQ, it is optimal to use the total transmission power
during the first transmission attempt.

\textbf{\emph{Practical Aspects}}: For limited real-time computational
resources, one can solve the optimization problems offline by using
nonlinear optimization techniques and store the results in a lookup
table. For real-time online power allocation, using nonlinear optimization
techniques to solve (\ref{eq:Opt_prob_1}) may be too costly. In such
cases, to achieve low target rate-outage probability, the simple closed-form
`PPA-GPP approach' is more computationally efficient and can provide
closer performance to the \textquoteleft PPA-exact method\textquoteright .
If high outage probability (e.g., around $10^{-3}$ or higher) is
acceptable, then one can use the EPA method.

\section{Conclusions\label{sec:Conclusions}}

We considered the problem of energy-efficient adaptive power allocation
in IMIMO systems. In general, these optimization are difficult to
solve as the rate-outage probability expressions are not mathematically
tractable. We developed methods to convert the rate-outage probability
expressions into a tractable form and solved the optimization problems
using interior-point algorithms. We used asymptotically equivalent
expressions of rate-outage probability and presented an unified geometric
programming formulation for which the closed-form solution is derived.
Possible extensions to the current work include, i) considering IMIMO
systems with a subset of antennas transmitting (as in generalized
spatial modulation) in each ARQ round instead of a single transmit
antenna; ii) solving the optimization problems with the objective
of minimizing the average delay or maximizing the long-term average
throughput.

\appendices{ }

\section{Proof of $q_{i}\left(t\right),2\leq i\leq l$}

Define $Q_{i}\left(s\right)$ as in \cite{IMIMO}:
\begin{align}
Q_{i}\left(s\right) & =\frac{U\left(N,N+s,\frac{1}{P_{i}}\right)}{P_{i}^{N}}\nonumber \\
 & =\int_{0}^{\infty}\frac{r^{N-1}\left(r+1\right)^{s-1}e^{-\frac{r}{P_{i}}}}{P_{i}^{N}\Gamma\left(N\right)}dr,i=2,\cdots,l\label{eq:G_i_def}
\end{align}
where $U\left(a,b,c\right)$ is the Tricomi confluent hypergeometric
function \cite{Tables_Book}, and $q_{i}\left(t\right),i=2,\cdots,l$
can be obtained as:
\begin{align}
q_{i}\left(t\right) & =\!L^{-1}\left(Q_{i}\left(s\right)\right)\nonumber \\
 & =\int_{0}^{\infty}\!\underbrace{\left[\!\frac{1}{2\pi j}\int_{c-j\infty}^{c+j\infty}\!e^{st}\left(r+1\right)^{s-1}ds\!\right]}_{=\mathcal{L}^{-1}\left(\left(r+1\right)^{s-1}\right)}\!\frac{r^{N-1}e^{-\frac{r}{P_{i}}}}{P_{i}^{N}\Gamma\left(N\right)}dr\!\nonumber \\
 & =\!\frac{e^{t}\!\left(e^{-t}-1\right)^{N-1}e^{\frac{1-e^{-t}}{P_{i}}}}{P_{i}^{N}\Gamma\left(N\right)}u\left(\!-t\right)\label{eq:Laplace_inverse}
\end{align}
In (\ref{eq:Laplace_inverse}), we used the relation $\mathcal{L}^{-1}\left(\left(r+1\right)^{s-1}\right)=\frac{1}{r+1}\mathcal{L}^{-1}\left(\left(r+1\right)^{s}\right)=\frac{1}{r+1}\mathcal{L}^{-1}\left(e^{s\ln\left(r+1\right)}\right)=\frac{1}{r+1}\delta\left(t+\ln\left(r+1\right)\right)$
and $\delta\left(.\right)$ is the Dirac delta function.

\section{Proof of Theorem 1}

We write the Lagrangian function of (\ref{eq:Opt_prob_IMIMO_GPP})
as:
\begin{align}
\mathscr{L}\left(P_{1},\cdots,P_{L},\lambda,\mu_{1},\cdots,\mu_{L}\right) & =\frac{W_{L}}{\left(P_{1}^{N}\cdots P_{L}^{N}\right)}\!+\nonumber \\
 & \!\!\!\!\!\!\!\!\!\!\!\!\!\!\!\!\!\!\!\!\!\!\!\!\!\!\!\!\!\!\!\!\!\!\!\!\!\!\!\!\!\!\!\!\!\!\!\!\!\!\!\!\!\!\!\!\!\!\!\!\!\!\!\!\!\!\!\!\!\!\lambda\!\left(\!P_{1}\!+\!\sum_{l=2}^{L}\!P_{l}\frac{W_{l-1}}{\left(P_{1}^{N}\cdots P_{l-1}^{N}\right)}\!-\!\overline{E}_{\mathrm{given}}\!\right)-\sum_{l=1}^{L}\mu_{l}P_{l}\label{eq:Lagrangian}
\end{align}

where $\lambda,\mu_{1},\cdots,\mu_{L}$ are the Lagrangian coefficients.
Since the Karush-Khun-Tucker (KKT) conditions are necessary for an
optimal solution, we have:\begin{subequations}
\begin{align}
\!\!\!\!\!\!\!\!\!\!\!\!\!\!\!\!\!\!\!\!\!\!\!\!\!\!\!\!\!\!\left.\frac{\partial\mathscr{L}}{\partial P_{l}}\right|_{\left(P_{1}^{*},\cdots,P_{L}^{*},\lambda^{*},\mu_{1}^{*},\cdots,\mu_{L}^{*}\right)}=0,\mbox{ for }1\leq l\leq L\label{eq:KKT_1}\\
\lambda^{*}\!\left(P_{1}^{*}\!+\!\sum_{l=2}^{L}\!P_{l}^{*}\!\frac{W_{l-1}}{\left(P_{1}^{*}\cdots P_{l-1}^{*}\right)^{N}}-\overline{E}_{\mathrm{given}}\right)=0\label{eq:kkt_2}\\
\mu_{l}^{*}P_{l}^{*}=0,\mbox{ for }1\leq l\leq L\label{eq:kkt3}
\end{align}
\end{subequations}

Since the objective is to minimize $W_{L}/\left(P_{1}^{N}\cdots P_{L}^{N}\right)$,
from (\ref{eq:kkt3}), it is clear that $\mu_{l}^{*}=0,\mbox{ for }1\leq l\leq L$.
Considering (\ref{eq:KKT_1}) for $l=L$ and simplifying, we have
$\lambda^{*}=NW_{L}/\left(W_{L-1}\left(P_{L}^{*}\right)^{N+1}\right)$.
Now considering (\ref{eq:KKT_1}) $\mbox{for }1\leq l\leq L-1$, together
with $\lambda^{*}$, we obtain
\begin{align}
\left(P_{l}^{*}\right)^{N+1} & =\frac{1}{W_{l-1}\prod_{m=l+1}^{L}\!\left(\!P_{m}^{*}\right)^{N}}\left(\!\vphantom{\left[\prod_{m=k+1}^{L}\left(P_{m}^{*}\right)^{N}\right]}\left(N+1\right)W_{L-1}\left(\!P_{L}^{*}\!\right)^{N+1}\!+\!\right.\nonumber \\
 & \left.N\sum_{k=l}^{L-2}W_{k}P_{k+1}^{*}\left[\prod_{m=k+1}^{L}\left(P_{m}^{*}\right)^{N}\right]\right).\label{eq:recursive_1}
\end{align}
Substituting for $\left(P_{l+1}^{*}\right)^{N+1}$ in (\ref{eq:recursive_1})
and simplifying, we obtain the recursive relation for $2\leq l\leq L$
as:
\begin{equation}
P_{l}^{*}=\frac{W_{l-2}}{\left(1+N\right)W_{l-1}}\left(P_{l-1}^{*}\right)^{N+1},\label{eq:recursive_2}
\end{equation}
Now to solve for $P_{1}^{*}$, we use the fact that the average energy
constraint should be satisfied with equality at the optimal solution,
i.e.,
\begin{equation}
P_{1}^{*}+\frac{W_{1}P_{2}^{*}}{\left(P_{1}^{*}\right)^{N}}+\frac{W_{2}P_{3}^{*}}{\left(P_{1}^{*}P_{2}^{*}\right)^{N}}+\frac{W_{L-1}P_{L}^{*}}{\left(P_{1}^{*}\cdots P_{L-1}^{*}\right)^{N}}=\overline{E}_{\mathrm{given}},\label{eq:avg_power_constraint}
\end{equation}
Now using (\ref{eq:recursive_2}) recursively in (\ref{eq:avg_power_constraint})
with $W_{0}=1$, we obtain the solution for $P_{1}^{*}$ as in (\ref{eq:Optimal_P_1}).

\pagebreak{}
\end{document}